\documentstyle[12pt]{article}
\begin{document}
\begin{titlepage}
\title{The Polynomial Formulation of the $U(1)$ Non-Linear $\sigma$-Model
in 2 Dimensions}
\author{C. D. Fosco and T. Matsuyama
\footnotemark[1]
\\ \\ University of Oxford\\
Department of Physics, Theoretical Physics\\
1 Keble Road, Oxford OX1 3NP, UK }
\vspace{1cm}

\baselineskip=21.5pt
\begin{abstract}
\vskip 2cm

\vskip 0.7cm

\footnotetext[1]{Permanent Address: Department of Physics, Nara University of
 Education, Takabatake-cho, Nara 630, JAPAN}

\end{abstract}
\maketitle
\end{titlepage}
\baselineskip=21.5pt
\parskip=3pt

The $SU(N)$ non-linear $\sigma$-model~\cite{sigma} dynamics can be
described by using a `polynomial' action which is first order in the
derivatives of the fields~\cite{tow,car,ces}.  The basic idea is that the
usual action may also be written in terms of
a ($SU(N)$) 1-form field $L_{\mu}$ which satisfies the Maureer-Cartan
(flatness) equation $F_{\mu \nu}(L) = 0$~\cite{slav}.
The polynomial formulation consists in writing the original action
in terms of the vector field, adding a Lagrange multiplier term
which enforces the Maureer-Cartan equation. As $F_{\mu \nu}$ is a
2-form field (i.e., an antisymmetric tensor), so must be the Lagrange
multiplier.

 From the usual description based on fields with values on a non-linear group
manifold, and with the proper group-invariant path integral measure, one goes
to another one in terms of a non-Abelian vector gauge field plus an
antisymmetric (2-form) tensor field, both with linear path integral
measures~\cite{car}. In three and higher spacetime dimensions there appears
a gauge symmetry under transformations of the antisymmetric tensor
field~\cite{anti} which, when gauge fixed, requires of course the
introduction of the corresponding Faddeev-Popov ghosts.
In four dimensions, this symmetry is reducible~\cite{teit}, thus a proper BRST
treatment prescribes the introduction of ghosts of ghosts.
Note also that the bosonic part of the action is first order in the
derivatives, and, depending on the gauge fixing one uses, the ghosts'
action may or may not be first order.

The $O(2)$ ($\sim U(1)$) non-linear $\sigma$-model in two (Euclidean)
dimensions (or $X-Y$ model) is the simplest possible example of application
for the polynomial formulation.
Besides the non-existence of the gauge symmetry linked to the
antisymmetric tensor field $\theta_{\mu \nu}$ in higher dimensions,
this field can be decomposed as: $\theta_{\mu \nu} = \epsilon_{\mu
\nu} \, \theta$ where $\theta$ is a pseudoscalar field.
Despite these simplifications, the model itself is far from trivial,
and in its Statistical Mechanics version it undergoes the celebrated
Kosterlitz-Thouless phase transition~\cite{kost}. Indeed, this phase
transition occurs precisely because the system has room for the existence
of vortices.
It is also well known that it can be mapped to a Sine-Gordon
model~\cite{zinn}, the correspondence between the $O(2)$ spin field and
the Sine-Gordon field being non-local. Thus the question presents itself
how does one introduce vortices or any other singular configuration.

The object of this paper is to study the polynomial formulation
for the $O(2)$ model in two dimensions, showing how some familiar results of
the usual formulation reemerge in a simpler way, and also how some
extensions can be implemented. In particular, the
Sine-Gordon description is obtained simply by integrating out the
vector field and thus deriving an `effective' action for the Lagrange
multiplier, which becomes precisely equal to the Sine-Gordon one.

The structure of this paper is as follows: In section 1 we briefly
review the model in its usual formulation, describing some properties
which we reformulate in their polynomial form in section 2.
In section 3 we construct the mapping to the Sine-Gordon
model, and in section 4 we show how to introduce strings of vortices.

\section{The non-polynomial or `second order' formulation}
The model is usually defined in terms of the Euclidean action (we
follow the presentation of ref.~\cite{zinn})
\begin{equation}
S \,=\, \frac{1}{2 t} \, \int d^2 x \,\, \partial_{\mu} {\bf s}(x)
\cdot \partial_{\mu} {\bf s}(x) \;
\end{equation}
where ${\bf s}(x)$ is a two-component, real, continuum spin field
\begin{equation}
{\bf s}(x) \,=\, ( s_1 (x) , s_2 (x) ) \,\, ,
\,\, {\bf s}^2 (x) = 1 \; ,
\label{002}
\end{equation}
and $t$ is a (dimensionless) parameter which plays the role of
a temperature.

To solve the constraint on the modulus of ${\bf s}$ one can parametrize
it as
\begin{equation}
{\bf s}(x) \,=\, (\cos \phi (x) , \sin \phi (x) )
\label{003}
\end{equation}
where now $\phi$ must be pseudoscalar under parity transformations
of the two-dimensional spacetime (reflections about one of the axis),
and one should note that physical quantities are $2 \pi$-periodic functions of
$\phi$ (i.e., it is an angular variable). The $O(2)$ symmetry has
been transformed into invariance under rigid translations of $\phi$.
The partition function is then
\begin{equation}
{\cal Z} \,=\, \int {\cal D} {\bf \phi} \, \exp [- S(\phi) ] \;,
\label{004}
\end{equation}
where
\begin{equation}
S(\phi) \,=\, \int d^2 x \,\,\frac{1}{2 t} \, \partial_{\mu} {\bf \phi}(x)
\partial_{\mu} {\bf \phi}(x) \;.
\label{005}
\end{equation}
Every correlation function of spin variables can be obtained by
linear combination of correlation functions of exponentials of
the field $\phi$
\begin{equation}
\langle \,\prod_{j = 1}^n \exp [ i \epsilon_j \phi (x_j) ] \,\rangle \;,
\; \epsilon_j = \pm 1 \, , \, \forall j \,
\end{equation}
where the $x_j$'s are the arguments of the spin fields in the corresponding
correlation function.
In the spin-wave approximation one neglects singular configurations
related to the periodicity of $\phi$, and then the correlation function
(6) can be exactly calculated as
\begin{equation}
\langle \, \prod_{j = 1}^n \exp [ i \epsilon_j \phi (x_j) ] \,\rangle \,
= \, {(\frac{\Lambda}{m})}^{- \frac{t n}{4 \pi}} \,
\prod_{j < k}^n \, {(m \mid x_j - x_k \mid)}^{\frac{t}{2 \pi} \epsilon_j
\epsilon_k} \; ,
\label{007}
\end{equation}
where $\Lambda$ and $m$ are UV and IR cutoffs respectively. When one
takes the $m \to 0$ limit, the only non-zero correlation functions
are the ones which satisfy the `neutrality' condition:
\begin{equation}
\sum_{j = 1}^n \epsilon_j \,=\, 0 \;,
\label{008}
\end{equation}
and this is precisely the condition for $O(2)$-invariance of the correlation
functions; i.e., invariance under translations of $\phi$. In particular
this implies the vanishing of the average of the one-spin function
(as prescribed by the Mermin-Wagner theorem~\cite{merm}) and,
regarding the 2-point correlation function, (\ref{007}) yields
\begin{equation}
\langle {\bf s}(x) \cdot {\bf s}(y) \rangle \,=\,
{(\frac{\Lambda}{m})}^{-\frac{t}{2 \pi}} {( m \mid x - y \mid)}^{-
\frac{t}{2 \pi}} \;.
\label{009}
\end{equation}
So far the discussion has been confined to the low-temperature phase,
characterized by an algebraic decaying of the 2-point correlation
function (\ref{009}). Let us consider now the high-temperature phase,
where vortices can appear unconfined, and the periodicity of $\phi$
is important. The vortices are finite-energy configurations
such that the angular field $\phi$ changes by  $2 \pi n $, where $n$
is an integer, when one moves around a point, the `center' of the vortex.
$n$ is known as the `winding' of the vortex.
A typical N-vortex configuration, with centers $X_j$,  (j=1, ..., N),
and windings ${n_j}$ is
\begin{eqnarray}
\phi_N (x, \{X_j, n_j \}) &=& \sum_{j =1}^N \, n_j \, \phi_j (x,{X_j})
\nonumber\\
\phi_j (x,X_j) &=& \arctan [\frac{{(x - X_j)}_2}{{(x - X_j)}_1}]
\;.
\label{010}
\end{eqnarray}
The contribution of ({\ref{010}}) to the action is then easily evaluated
\begin{equation}
S_N \,=\, -\frac{\pi}{t} \sum_{j = 1}^N n_j^2 \log (\frac{\Lambda}{m}) \,
- \, \frac{2 \pi}{t} \, \sum_{j<k}^N n_j n_k \log \mid X_j - X_k \mid \;.
\label{011}
\end{equation}
In the last equation, the neutrality condition for the vortices'
charges must also be assumed in order to have a non-zero action when
the IR cutoff is removed. When also spin waves are present, the
total action becomes the sum of (\ref{011}) and the usual spin
wave part.
In what follows the (usual) approximation is made of
considering only windings equal to $\pm 1$ for the vortices, since they
are the most relevant. Due to this constraint, each sector must contain a
fixed number of vortex-antivortex pairs, and we shall include a
fugacity $\eta$ plus a combinatorial (classical) factor to take into account
the indistinguishability of the vortices of equal charge.
The partition function is then calculated by summing (with the appropriate
weight factors) over all the possible configurations within each
topological sector, and then over all the topological sectors.
Sumarising,
\begin{equation}
{\cal Z} \,=\, \sum_{N = 0}^{\infty} \frac{{\eta}^{2 N}}{{(N !)}^2}
\int \prod_{j =1}^N d X^2_j d Y^2_j \, {\cal Z}_N (\{ X_j , Y_j \}) \;,
\label{012}
\end{equation}
where ${\cal Z}_N$ is the partition function for the spin variables in
the presence of $N$ vortex-antivortex pairs, with coordinates
$X_j, Y_j$, respectively. One easily realises
that the spin-wave contribution factors out, and the total partition
function becomes the product of the partition function of a free
scalar field by a Coulomb gas partition function. This Coulomb gas
partition function can then be mapped to a Sine-Gordon partition
function~\cite{zinn}. However, the correspondence between the {\em fields}
in both partition functions is somewhat {\em ad hoc} in this framework.

\section{The polynomial or `first order' formulation}
We know from the previous section that the only non-trivial
correlation functions are the $O(2)$-invariant ones, and this is
tantamount of invariance under translations in $\phi$. It seems then
natural to look for the possibility of describing the system completely
in $O(2)$ invariant terms. We realize that the simplest possible
$\phi$-translation invariant field variable is
\begin{equation}
L_{\mu} (x) \,=\, \frac{1}{g} \partial_{\mu} \phi (x) \;,
\label{013}
\end{equation}
where $g$ is a constant with dimensions of mass, introduced to
make $L_{\mu}$ dimensionless. We may of course rewrite the action
(\ref{005}) in terms of $L_{\mu}$ only, but we must also take into
account the fact that $L_{\mu}$ is not entirely arbitrary, but
a `pure gauge' field, i.e.,
\begin{equation}
\epsilon_{\mu \nu} \partial_{\mu} L_{\nu} (x) \,=\, 0 \;.
\label{014}
\end{equation}
When vortices are present, (\ref{014}) is relaxed, allowing for
a discrete set of points where the rhs is non-zero.

We construct the first order action for the system in the spin wave
sector by rewriting (\ref{005}) in terms of $L_{\mu}$ and then
adding a Lagrange multiplier term for the condition (\ref{014}):
\begin{equation}
S_{sw} \,=\, \int d^2 x ( \frac{g^2}{2 t} L_{\mu} L_{\nu} -
i g \theta \epsilon_{\mu \nu} \partial_{\mu} L_{\nu} )
\label{015}
\end{equation}
where $\theta$ is a scalar field which enforces condition
(\ref{014}). $O(2)$ invariant correlation functions are then
calculated in terms of $L_{\mu}$, observing that~\cite{ces}
\begin{equation}
\exp \{ i [ \phi (x_1) - \phi (x_2)] \} \,=\, \exp [ i g \int_{x_2}^{x_1}
d z_{\mu} L_{\mu} (z) ]
\label{016}
\end{equation}
where the line integral is taken along any smooth path joining $x_2$ to
$x_1$. (\ref{016})
{\em defines} the lhs in terms of the rhs\footnote{This is done just
to make contact with the usual description, but there is no
necessity to introduce the spin fields, neither to solve the
constraints on $L_{\mu}$  within the polynomial
formulation.}. In order to give a consistent definition of a local
spin field, we should require (\ref{016}) to be path-independent.
This will happen as long as:
\begin{equation}
g \oint_{C} d z_{\mu} L_{\mu} (z) \,=\, 2 \pi N
\label{017}
\end{equation}
(where $N$ is any integer) for every closed curve $C$.

The general non-zero correlation function
of the usual formulation will satisfy condition (\ref{008}), then
for each $\epsilon_j = +1$ in (\ref{007}), there must
be an $\epsilon_k = -1$. Whence the general correlation function of
the usual formulation
can be constructed by forming `neutral pairs' like the lhs of (\ref{016}),
and then  writing them in terms of the corresponding Wilson line on
the rhs (of course there are many different ways to chose the pairings,
all giving the same result).

The generalization
of (\ref{014}) to the case when $N$ vortex-antivortex pairs are present
is just
\begin{eqnarray}
\epsilon_{\mu \nu} \partial_{\mu} L_{\nu} (x) \,&=&\, \rho_N (x) \nonumber
\\
\rho_{N} (x) &=& \frac{2 \pi}{g} \sum_{j = 1}^{N} [\delta (x - X_j)
- \delta (x - Y_j) ] \; ,
\label{018}
\end{eqnarray}
where we use the notation of Equation (12).
Within the topological sector defined by (\ref{018}), the generating
functional of $L_{\mu}$ correlation functions is
\begin{equation}
{\cal Z}_N (J) \,=\, \int {\cal D}L_{\mu} \, {\cal D} \theta \,\,
\exp ( - S_N + \int d^2 x J_{\mu} L_{\mu} ) \; ,
\label{019}
\end{equation}
where
\begin{equation}
S_N \,=\, \int d^2 x \,\{ \frac{g^2}{2 t} L_{\mu}(x) L_{\mu}(x) \,-\, i g
\theta \,[\epsilon_{\mu \nu} \partial_{\mu} L_{\nu} - \rho_N (x)]\,
\}
\label{020}
\end{equation}
with $\rho_N$ as defined by (\ref{018}). Let us calculate then the
spin-spin correlation function in this sector
\begin{equation}
\langle {\bf s}(x) \cdot {\bf s}(y) \rangle_N \,=\, \Re {\int
{\cal D}L_{\mu} {\cal D}\theta \, \exp [i \int_{y}^{x} d y_{\mu}
L_{\mu} (y) ] \, \exp (- S_N) }\;
\end{equation}
(where $\Re$ means real part of).
Thus (21) can be obtained from ${\cal Z}_N(J)$ just by specifying
the current $J_{\mu}$ which reproduces the `Wilson line' for $L_{\mu}$,
i.e.,
\begin{equation}
J_{\mu} (u) \,=\, i g \int_0^1 ds \, \frac{d z_{\mu}}{d s} \, \delta
(u - z(s)) \;.
\label{022}
\end{equation}
We first calculate ${\cal Z}_N (J)$ for arbitrary $J$ and then we take it to
be equal to (\ref{022}).
Integrating  $L_{\mu}$ in (\ref{019}), we get
\begin{equation}
{\cal Z}_N (J) \,=\, \int {\cal D} \theta \, \exp [ - \frac{t}{2}
\partial_{\mu} \theta \partial_{\mu} \theta - \frac{t}{2 g^2}
J^2 + \theta ( - \frac{t}{g} \epsilon_{\mu \nu} \partial_{\mu}
J_{\nu} + i g \rho_N ) ] \;.
\label{023}
\end{equation}
The integral over $\theta$ is also Gaussian. The final result is
\begin{eqnarray}
{\cal Z}_N (J)\,&=&\, \exp [-\frac{t}{2 g^2} \int  d^2 x \partial \cdot
J \partial^{-2} \partial \cdot J + \frac{g^2}{2 t} \int d^2 x
\rho_N \partial^{-2} \rho_N \nonumber \\
&-&  \int d^2 x
\epsilon_{\mu \nu} \partial_{\mu} J_{\nu} \partial^{-2} \rho] \;.
\label{024}
\end{eqnarray}
Using for $J_{\mu}$ the explicit form (\ref{022}), (\ref{024}) yields
\begin{eqnarray}
\langle {\bf s}(x) \cdot {\bf s}(y) \rangle_{N}
& = & \exp [ -\frac{t}{2 \pi} \log \mid x - y \mid ]
\exp [ -\frac{\pi}{t} \sum_{i,j=1}^N \log \mid X_i - Y_j \mid ]
\nonumber\\
& & \times \cos \{ \sum_{k =1}^N [\alpha (x - X_k) -\alpha (x -Y_k)
- \alpha (y - X_k) + \alpha (y - Y_k)  ] \} \nonumber\\
& \equiv & G(x , y ; \{ X_k , Y_k \}; t ) \,
\end{eqnarray}
where we have absorbed the divergent factors in a renormalization
of the fields, and we use the function $G$ to denote explicitly the
dependence of the correlation function also on the coordinates of the
vortices and the temperature.

Now we particularize equation (25) for the case $N = 1$
\begin{eqnarray}
G (x , y; X, Y; t) &=&
\exp[ -\frac{t}{2 \pi} \log \mid x - y \mid ] \,
\exp [ -\frac{2 \pi}{t}  \log \mid X - Y \mid ]
\nonumber\\
&\times& \cos[
\alpha (x - X) - \alpha (x - Y)
- \alpha (y - X) + \alpha (y - Y) ] ,
\label{aux}
\end{eqnarray}
where $\alpha (x) = \arctan x_2/x_1$ for any $x = (x_1,x_2)$.
The correlation function (\ref{aux}) is symmetric under the
interchange of spin and vortex coordinates, plus a transformation
of the temperature:
\begin{equation}
x \leftrightarrow X \;,\; y \leftrightarrow Y \;\; ; \;\; t \to \frac{4
 \pi^2}{t} \; .
\label{d1}
\end{equation}
Of course, this symmetry is inherent to the model, and not
a consequence of using the polynomial formulation. However,
we will need it to realise this symmetry as the result of some
invariance of the path integral under a
transformation of the {\em fields}. To achieve this, we need a description where
spins field and vortices field appear in a
more symmetrical way.
$L_{\mu}$ is the spin field and one can realize that $\theta$ plays the role of
 a vortex field.
Indeed, the path integral representation of (\ref{aux}) as defined by (21) is
\begin{eqnarray}
\langle {\bf s} (x) \cdot {\bf s} (y) \rangle
&=& \Re \int {\cal D} L_{\mu} \, {\cal D} \theta \,
\exp [ \int d^2 x ( -\frac{g^2}{2 t} L_{\mu} L_{\mu}
+ i g \theta \epsilon_{\mu \nu} \partial_{\mu}  L_{\nu} )] \nonumber\\
&\times&\exp (i g \int_y^x d z_{\mu} L_{\mu})
\, \exp [- 2 \pi i (\theta (X) - \theta (Y))]
\;,
\label{d2}
\end{eqnarray}
where one can see that the exponential of $\theta$ creates vortices when
 averaged with the spin wave action.
We note that
\begin{eqnarray}
& & \int {\cal D} \theta \exp [\int d^2 x i g \theta \epsilon_{\mu \nu}
\partial_{\mu} L_{\nu}]
\exp [ - 2 \pi i ( \theta (X) - \theta (Y) ) ] \nonumber\\
&=& \int {\cal D} \theta_{\mu} {\cal D} \Lambda
\exp [ \int d^2 x ( -i g^2 \epsilon_{\mu \nu} L_{\mu}
\theta_{\nu} + i g \Lambda \epsilon_{\mu \nu} \partial_{\mu}\theta_{\nu}
) ] \nonumber\\
& \times & \exp[ - 2 \pi i g \int_Y^X d z_{\mu} \theta_{\mu} (z) ] \; ,
\label{d3}
\end{eqnarray}
where $\Lambda$ is a new Lagrange multiplier field, which enforces
the condition $\epsilon_{\mu \nu} \partial_{\mu} \theta_{\nu} = 0$,
which is solved by $\theta_{\mu} (x) = \frac{1}{g} \partial_{\mu}
\theta (x)$.
Inserting (\ref{d3}) into (\ref{d2}) we get a more
symmetrical description, in terms of the spin field $L_{\mu}$ and vortex field
 $\theta_{\mu}$,
\begin{eqnarray}
\langle {\bf s} (x) \cdot {\bf s} (y) \rangle
& = & \Re \int {\cal D} L_{\mu} \, {\cal D} \theta_{\mu} \,
{\cal D} \Lambda \nonumber\\
&\times& \exp [ \int d^2 x ( -\frac{g^2}{2 t} L_{\mu} L_{\mu}
- i g^2 \epsilon_{\mu \nu} \theta_{\mu}  L_{\nu}
+ i g \Lambda \epsilon_{\mu \nu} \partial_{\mu} \theta_{\nu} )] \nonumber\\
& \times &
\exp (i g \int_y^x d z_{\mu} L_{\mu}(z))
\exp [- 2 \pi i g\int_Y^X d z_{\mu} \theta_{\mu}(z)]\;.
\label{d4}
\end{eqnarray}
Thus, the spin-spin correlation function corresponds to the average of
the product of two `Wilson lines', one for $L_{\mu}$ and the other for
$\theta_{\mu}$:
\begin{equation}
\langle {\bf s}(x) \cdot {\bf s} (y) \rangle \,=\,
\Re \langle \exp [ i g \int_y^x dz \cdot L] \, \exp [2 \pi  i g
\int_Y^X dz \cdot \theta ] \rangle \;,
\label{e1}
\end{equation}
where the average is performed with the action
\begin{equation}
S \,=\, \int d^2 x \{ \frac{g^2}{2 t} L^2 + i g^2 \epsilon_{\mu \nu}
\theta_{\mu} L_{\nu} - i g \Lambda \epsilon_{\mu \nu} \partial_{\mu}
\theta_{\nu} \} \;.
\label{e2}
\end{equation}
Then, the duality transformation amounts to performing the following
transformation in thew action (\ref{e2}):
\begin{equation}
L_{\mu} (x) \to 2 \pi \theta_{\mu} (x) \;,\; \theta_{\mu} (x) \to
\frac{1}{2 \pi} L_{\mu} (x) \;,\; t \to \frac{4 \pi^2}{t} \;.
\label{e3}
\end{equation}
One easily verifies that the average of the same Wilson's lines
operators with the transformed action gives the transformation
(\ref{d1}).
Note that the effect of this change of variables is to interchange the roles
of vortices and spins, as well as low and high temperatures.
In this sense, $t = \frac{4 \pi^2}{t}$ is a kind of self-dual point,
where spins and vortices are interchangeable.

\section{Mapping to the Sine-Gordon Model.}
The total partition function is defined following (\ref{012})
\begin{equation}
{\cal Z} \,=\, \sum_{N = 0}^{\infty} \frac{1}{ {(N!)}^2}  \eta^{2 N}
\int \prod_{j = 1}^N d^2 X_j d^2 Y_j {\cal Z}_{N} (0) \;,
\label{026}
\end{equation}
with
\begin{eqnarray}
{\cal Z}_{N}(0) (\{X_j , Y_j \}) &=& \int {\cal D} L_{\mu} {\cal D}
\theta \,\, \exp \{ - \int d^2 x [\frac{g^2}{2 t} L_{\mu} L_{\mu}
- i g \theta (\epsilon_{\mu \nu} \partial_{\mu} L_{\nu} - \rho_{N})] \}
\; ,\nonumber\\
\rho_{N} (x) &=& \frac{2 \pi}{g} \sum_{j = 1}^{N} \, [\delta (x - X_j)
- \delta (x - Y_j) ] \; .
\label{027}
\end{eqnarray}

Performing the Gaussian integration over $L_{\mu}$ in (\ref{027}), and
using the delta functions in the definition of $\rho_{N}$, we can
rewrite ${\cal Z}_{N}$ as
\begin{equation}
{\cal Z}_{N} \,=\, \langle \, \prod_{j=1}^N \exp [ 2 \pi i ( \theta
(X_j) - \theta (Y_j)) ] \, \rangle \;,
\label{028}
\end{equation}
where the average $\langle \rangle$ is defined by
\begin{eqnarray}
\langle F(\theta) \rangle &=& \int {\cal D} \theta
F(\theta) \, \exp [ - S(\theta) ] \; , \nonumber \\
S(\theta) &=& \frac{t}{2} \int d^2 x \partial_{\mu} \theta
\partial_{\mu} \theta \;.
\label{029}
\end{eqnarray}
Using (\ref{029}) in (\ref{026}), we can rewrite the total partition
function ${\cal Z}$ as
\begin{equation}
{\cal Z} \,=\, \sum_{N = 0}^{\infty} \frac{\eta^{2 N}}{{(N !)}^2}
\langle \, {[ \int d^2 x e^{- 2 \pi i \theta (x)} ]}^N
{[ \int d^2 y e^{+ 2 \pi i \theta (y)} ]}^N  \, \rangle \;.
\label{030}
\end{equation}
Taking now into account the fact that only products of exponentials that
satisfy the `neutrality' condition are non-zero, it is straightforward
to check that
\begin{equation}
{\cal Z} \,=\, \langle \, \exp ( 2 \eta \int d^2 x \cos ( 2 \pi \theta ) )
\, \rangle
\label{/31}
\end{equation}
or
\begin{equation}
{\cal Z} \,=\, \int {\cal D} \theta \, \exp \{ - \int d^2 x [ \frac{t}{2}
{(\partial \theta)}^2 - 2 \eta \cos (2 \pi \theta )] \}
\label{032}
\end{equation}
which is the desired Sine-Gordon action in terms of the Lagrange
multiplier field $\theta$. This coincides with the result obtained
by more traditional methods.

The effective action for $\theta$ in (\ref{032}) was obtained under
the assumption that the vortices can only have charges equal to $\pm 1$.
Let us consider now an extension of the model. It consists of
relaxing this constraint, allowing for the charges to be equal
to plus or minus any real number $q$. Of course, the local spin
interpretation will no longer be true, since the definition (\ref{016})
requires condition (\ref{017}). In this sense this can be considered as
an extension of the $O(2)$ model, which allows for vortices of
non-integer charge. However, we will first assume a unique value for $q$
in the model.
The topological sectors will then have $2N$ vortex-antivortex
pairs as before.
Again only neutral combinations will have a finite weight.
Of course we will also average over the positions of the vortices,
and include the corresponding combinatorial factors.
One can verify that instead of obtaining the Sine-Gordon partition
function (\ref{032}), we get
\begin{equation}
{\cal Z} \,=\, \int {\cal D} \theta \, \exp \{ - \int d^2 x [ \frac{t}{2}
{(\partial \theta)}^2 - 2 \eta \cos (2 \pi q \theta )] \} \;.
\label{f1}
\end{equation}
Thus, even when the change performed on the spin system is drastic,
the Sine-Gordon parameters change smoothly from $q=1$ to any $q$.
Things change more dramatically if we average now over all the
possible values of $q$, since this produces delta-functions of the
field $\theta$. The result is a kind of field-theoretic delta-function
model:
\begin{equation}
{\cal Z} \,=\, \int {\cal D} \theta \, \exp \{ - \int d^2 x [ \frac{t}{2}
{(\partial \theta)}^2 -  \eta \delta ( \theta )] \} \;.
\label{f2}
\end{equation}

\section{Strings of vortices}
We have shown how the usual point-like singularities (vortices)
are introduced in the polynomial formulation. Let us consider now
string-like singularities (which could be regarded as strings of
vortices). The obvious generalisation  of the procedure we followed
for the point-like case is to impose on $L_{\mu}$ a constraint like:
\begin{equation}
\epsilon_{\mu \nu} \partial_{\mu} L_{\nu} (x) \,=\, \rho_{st} (x) \;,
\label{131}
\end{equation}
where $\rho_{st}$ is the density appropriate to a string:
\begin{equation}
\rho_{st} (x) \,=\, \frac{2 \pi}{g} \int_0^1 d s \, q (s) \,
\delta [x - \gamma (s)]
\label{132}
\end{equation}
where the string's path is parametrized by $\gamma (s) : [0,1]
\to {\rm R}^2$, and $q(s)$ measures the density of vorticity along
the curve.
Of course we can consider more than one string, just by using in
the rhs of (\ref{132}) the sum of the densities corresponding to
the paths.
One can see that the spin configuration which corresponds
to a state like the one defined by (\ref{131}) and (\ref{132})
is:
\begin{eqnarray}
{\bf s}(x) &=& ( \, \cos \phi (x) , \sin \phi (x) \, ) \;,\nonumber\\
\phi (x) &=& \int_0^1 d s q(s) \arctan (x - \gamma (s)) \; .
\label{033}
\end{eqnarray}
As for the vortices, the non-local definition of the spin field will
be consistent only if condition (\ref{017}) is met for any curve\footnote{We
do not allow the curves to intercept any singularity.}.
When strings are present this implies:
\begin{equation}
\int_0^1 ds \,\, q_j(s) \,=\, N_j \;\;,\;\; \forall j \;,
\label{034}
\end{equation}
where the $N_j$'s are integers.
Also, in order for the action to be non-zero when the IR cut-off is
removed, a neutrality condition must be satisfied
\begin{equation}
\sum_{j=1}^N \int_0^1 \, q_j(s) \,=\, 0 \;.
\label{035}
\end{equation}
When only one string is present (\ref{035}) supersedes (\ref{034}), and thus
there is not true singularity since (\ref{033}) implies that the net rotation
of the spin in the local frame is null. Note that there is
a crucial difference between open and closed strings, because while in
the former the Wilson line definition works globally, in the latter it
applies only to one of the simply-connected regions into which the
space becomes divided (we recall that we cannot cross a singularity
with the Wilson line). Let us estimate the
Boltzmann weight of a configuration of a $N$ string-antistring
configuration.
It is straightforward to calculate the action due to this configuration
in the polynomial formulation. It becomes just the Coulomb energy of the
corresponding charge distribution:
\begin{equation}
S_{st} \,=\, \frac{\pi}{t} \sum_{i, j = 1}^N \int_0^1 ds_1 \int_0^1
ds_2 \,\, q_i (s_1) \,  q_j (s_2) \,\, \log \mid \alpha_i (s_1) -
\beta_j (s_2) \mid
\;,
\label{036}
\end{equation}
where $\alpha$ and $\beta$ parametrize the strings and antistrings
paths, respectively (we have not written the (divergent) self-energies).
With the natural definition of the mean position ($X_j$) of the
j-th string,
\begin{equation}
\log (x - X_j) \,=\, \int_0^1 ds q_j(s) \log (x - \alpha (s) ) \;
, \; ( \int_0^1 ds q_j (s) = + 1)  \;,
\label{037}
\end{equation}
and analogously for an antistring,
the action looks exactly like the one of $N$ vortex-antivortex pairs,
i.e.,
\begin{equation}
S_{st} \,=\, \frac{\pi}{t} \sum_{j,k=1}^N  \log \mid
X_j - Y_k \mid \;.
\label{038}
\end{equation}
Thus we have arrived at the conclusion that the Botzmann weight of
these configurations is like the one of the usual point-like vortices.
However, large strings  are suppressed in the partition function
because of the restriction about their positions. For example, the
volume inside a closed string cannot be occupied by another one, and
so they should be more strongly suppressed than the open ones.

\section*{Acknowledgements}
C. D. F. was supported by an European Community Postdoctoral Fellowship.
T. M. was supported in part by the British Council and the Daiwa
Anglo-Japanese Foundation. We also would like to express our
acknowledgement to Dr. I. J. R. Aitchison for his kind hospitality.


\begin{thebibliography}{100}
\bibitem{sigma} M. Gell-Mann and M. L\'evy, Nuovo Cimento {\bf 16},
705 (1960);
S. Weinberg, Phys. Rev. {\bf 166}, 1568 (1968);
J. Schwinger, Phys. Rev. {\bf 167}, 1432 (1968).
\bibitem{tow} D. J. Freedman and P. K. Townsed, Nucl. Phys.
B{\bf 177}, 282 (1981).
\bibitem{car}G. L. Demarco, C. D. Fosco and R. C. Trinchero,
Phys. Rev D{\bf 45}, 3701 (1992).
\bibitem{ces} C. D. Fosco and R. C. Trinchero, Phys. Lett. B{\bf 322},
97 (1994).
\bibitem{slav}A. A. Slavnov; Nucl. Phys. B{\bf 31}, 301 (1971).
\bibitem{anti}J. Thierry-Mieg, Nucl. Phys B{\bf 335}, 334 (1990);
W. Siegel, Phys. Lett. B{\bf 93}, 170 (1980).
\bibitem{teit} See for example: M. Henneaux and C. Teitelboim,
`Quantization of Gauge Systems', Princeton Univ. Press, p. 469 (1992).
\bibitem{kost} J. M. Kosterlitz and D. J. Thouless, J. Phys. C:
Solid State Physics {\bf 6} L97 (1973);
J. M. Kosterlitz, J. Phys. C: Solid State Physics {\bf 7} 1046 (1974);
J. V. Jos\'e, L. P. Kadanoff, S. Kirkpatrick and D. R. Nelson,
Phys. Rev. B{\bf 16} 1217 (1977).
\bibitem{zinn} J. Zinn-Justin, {\em Quantum Field Theory and
Critical Phenomena}, Oxford University Press, Second Edition (1993).
\bibitem{merm} N. D. Mermin and H. Wagner, Phys. Rev. Lett {\bf 17},
1133 (1966); S. Coleman, Comm. Math. Phys. {\bf 31}, 259 (1973).
\end{thebibliography}
\end{document}